\documentstyle[aps,floats,epsf,epsfig,fleqn,eqsecnum]{revtex}
\DeclareFontFamily{U}{msb}{}
\DeclareFontShape{U}{msb}{m}{n}{
<5><6><7><8><9> gen *msbm <10><10.95><12><14.4><17.28><20.74><24.88>msbm10}{}
\DeclareSymbolFont{AMSb}{U}{msb}{m}{n}
\DeclareMathSymbol{\GGG}{\mathbin}{AMSb}{'107}
\DeclareMathSymbol{\KKK}{\mathbin}{AMSb}{'113}
\DeclareMathSymbol{\MMM}{\mathbin}{AMSb}{'115}
\DeclareMathSymbol{\PPP}{\mathbin}{AMSb}{'120}
\DeclareMathSymbol{\RRR}{\mathbin}{AMSb}{'122}
\DeclareMathSymbol{\SSS}{\mathbin}{AMSb}{'123}

\newcommand{\phb}{\overline{\phi}}
\newcommand{\kb}{\overline{k}}

\newcommand{\XX}{{\bf x}}

\begin{document}
\draft
\setcounter{page}{0}

\title{
Threedimensional Local Porosity Analysis of Porous Media
}
\author{B. Biswa$\mbox{\rm l}^{1,2}$, C. Manwar$\mbox{\rm t}^{1}$ and R. Hilfe$\mbox{\rm r}^{1,3}$}
\address{
$\mbox{ }^1$ICA-1, Universit{\"a}t Stuttgart,
Pfaffenwaldring 27, 70569 Stuttgart\\
$\mbox{ }^2$Department of Physics \& Electronics,
Sri Venkateswara College, \\
University of Delhi, New Delhi - 110 021, India\\
$\mbox{ }^3$Institut f{\"u}r Physik,
Universit{\"a}t Mainz,
55099 Mainz, Germany}
\maketitle

\thispagestyle{empty}
\begin{abstract}
A quantitative comparison of the pore space
geometry for 
three natural sandstones is presented.
The comparison is based on local porosity theory
which provides a geometric characterization of
stochastic microstructures.
The characterization focusses on porosity and
connectivity fluctuations.
Porosity fluctuations are measured using local porosity
distributions while connectivity fluctuations are measured 
using local percolation probabilities.
We report the first measurement of local 
percolation probability functions for
experimentally obtained threedimensional 
pore space reconstructions.
Our results suggest to use
local porosity distributions and percolation
probabilities as a quantitative method to
compare microstructures between models 
and experiment.
\end{abstract}
\vspace{.7cm}
\begin{tabbing}
PACS: \= 61.43.G \hspace*{2ex}\= (Porous materials; structure),\\
\> 81.05.Rm \> (Porous materials; granular materials),\\
\> 47.55.Mh \> (Flows through porous media)
\end{tabbing}
{\em to appear in } Physica A

\newpage
\section{Introduction}

A large number of microscopic models have been proposed
to represent the microstructure of porous media 
\cite{fat56a,sch74,CD77,zim82,RS85,JA87,adl92,BT93,FJ95,JC96,AAMHSS97,OBA97}.
Representative microscopic models are a prerequisite for studying
transport properties such as fluid flow or sound propagation in 
oil reservoirs, aquifers or other random media.

Microscopic models are not unique, and hence it is 
necessary to have criteria for comparing them among each
other and to the experimental porous microstructure
\cite{hil92f,ABL94,hil95d,hil96g,sic97}.
This is particularly important for attempts to generate
porous microstructures in an automatic computerized 
process \cite{qui84,adl92,BT93,YT98}, or to decide
quantitatively whether the connectivity is percolationlike
as is often assumed in models \cite{OBFJMA91,BKS92,AAMHSS97}.

Detailed microscopic models contain so many
geometrical features that the rough comparison
based only on porosity and specific surface is 
insufficient.
The problem is to find general geometric characterization
methods to test how well a model 
represents the microstructure found in reality.
Given such tools they can then be used to constrain 
the input parameters of the models.

General geometric characterization methods traditionally
employ only porosities, specific surface areas,
and sometimes correlation functions \cite{sch74,ste85,dul92,adl92,BO96}.
Recently novel tools based on local porosity theory became
available for the comparison of stochastic microstructures
\cite{hil91d,hil92a,hil92b,hil92f,hil93b,hil93c,hil94b,hil95d}.
Local porosity theory is currently the most general geometric 
characterization method because it contains  as a special case 
also the characterization through correlation functions
(see \cite{hil95d} for details).

Local porosity theory contains two geometric
characteristics.
The first are local porosity distributions, the second
are local percolation probabilities \cite{hil95d}.
While local porosity distributions have been measured
previously on artificial and real samples 
\cite{hil92f,hil93c,hil94b,hil96c}
no reliable measurement has been made up to now for the
local percolation probabilities.
The main impediment has been the absence of accurate
threedimensional pore space representations for real rocks.

The objective of the work reported here has been to measure
the local percolation probabilities of natural sandstones
thereby providing a new geometric characteristic against
which microscopic pore space models can be compared.

\section{Measured Quantities}
\subsection{Local Porosity Distributions}
Local porosity distributions were originally introduced
as a quantitative substitute for pore size distributions 
\cite{hil91d}.
The idea is to measure porosity or other well defined
geometric observables within a bounded (compact) subset
of the porous medium and to collect these measurements
into various histograms (empirical probability densities).

Imagine a porous medium occupying a subset $\SSS\subset\RRR^d$ 
of the physical space ($d=3$ in the following).
For the data analysed here 
the set $\SSS$ is a rectangular parallelepiped whose
sidelengths are $M_1,M_2$ and $M_3$ in units of the 
lattice constant $a$ (resolution) of a simple cubic
lattice.
The sample $\SSS$ contains two disjoint subsets
$\SSS=\PPP\cup\MMM$ with $\PPP\cap\MMM=\emptyset$
where $\PPP$ is the pore space and $\MMM$ is the 
rock or mineral matrix and $\emptyset$ is the empty set.
In practice the sample is discretized, and the configuration
of the two sets $\PPP$ and $\MMM$ is given as an 
$M_1\times M_2\times M_3$-array of two numbers 
representing $\PPP$ and $\MMM$ respectively.
Let $\KKK(\XX,L)$ denote a cube of sidelength $L$
centered at the lattice vector $\XX$.
The set $\KKK(\XX,L)$ defines a measurement cell
inside of which local geometric properties such
as porosity or specific internal surface are
measured.
The local porosity in this measurement cell $\KKK(\XX,L)$ 
is defined as
\begin{equation}
\phi(\XX,L)=\frac{V(\PPP\cap\KKK(\XX,L))}{V(\KKK(\XX,L))}
\label{lpd1}
\end{equation}
where $V(\GGG)$ is the volume of the set $\GGG\subset\RRR^d$.
The local porosity distribution $\mu(\phi,L)$ is defined as
\begin{equation}
\mu(\phi,L) = \frac{1}{m}\sum_\XX\delta(\phi-\phi(\XX,L))
\label{lpd2}
\end{equation}
where $m$ 
is the number of placements of the measurement cell $\KKK(\XX,L)$.
The results presented below are obtained by placing $\KKK(\XX,L)$
on all lattice sites $\XX$ which are at least a distance $L/2$ 
from the boundary of $\SSS$, and hence in the following
\begin{equation}
m = \prod^3_{i=1}(M_i-L+1) 
\label{lpd3}
\end{equation}
will be used.
$\mu(\phi,L)$ is the empirical probability density function
(histogram) of local porosities.
Its support is the unit interval.

It is simple to determine $\mu(\phi,L)$ in the limits $L\rightarrow 0$
and $L\rightarrow \infty$ of small and large measurement cells.
For small cells one finds generally \cite{hil91d,hil95d}
\begin{equation}
\mu(\phi,L=0) = \phb\delta(\phi-1)+(1-\phb)\delta(\phi)
\label{lpd4}
\end{equation}
where 
\begin{equation}
\phb=V(\PPP\cap\SSS)/V(\SSS)
\label{bulk}
\end{equation}
is the bulk porosity.
If the sample is macroscopically homogeneous then
\begin{equation}
\mu(\phi,L\rightarrow\infty) = \delta(\phi-\phb)
\label{lpd5}
\end{equation}
indicating that in both limits the geometrical information
contained in $\mu(\phi,L)$ consists of the single number
$\phb$.
The macroscopic limit, however, involves the
question of macroscopic heterogeneity versus
macroscopic homogeneity (for more information 
see \cite{hil95d}).
In any case, if eqs. (\ref{lpd4}) and (\ref{lpd5}) hold
it follows that there exists a special length scale $L^*$ 
defined as
\begin{equation}
L^*=\min\{L : \mu(0,L)=\mu(1,L)=0\}
\label{lpd6}
\end{equation}
at which the $\delta$-distributions at $\phi=0$ and $\phi=1$ 
both vanish for the first time.

\subsection{Local Percolation Probabilities}
The local percolation probabilities characterize the connectivity
of measurement cells of a given local porosity.
Let 
\begin{equation}
\Lambda_\alpha(\XX,L)=\left\{
\begin{array}{r@{\quad:\quad}l}
1 & {\rm if~}\KKK(\XX,L){\rm ~percolates~in~``}\alpha\mbox{\rm
''-direction}\\[6pt]
0 & {\rm otherwise}
\end{array}
\right.
\label{lpp1}
\end{equation}
be an indicator for percolation. 
What is meant by ``$\alpha$''-direction is summarized in 
Table I.
\begin{center}
{\small TABLE I: Legend for index $\alpha$ of local percolation
probabilities $\lambda_\alpha(\phi,L)$.}\\[8pt]
\begin{tabular}{|c|c|} \hline
index $\alpha$ & meaning \\\hline
$x$ & $x$-direction\\
$y$ & $y$-direction\\
$z$ & $z$-direction\\
$3$ & ($x\wedge y\wedge z$)-direction\\
$c$ & ($x\vee y\vee z$)-direction\\
$0$ & ($\neg(x\vee y\vee z)$)-direction\\\hline
\end{tabular}
\end{center}
A cell $\KKK(\XX,L)$ is called ``percolating in the $x$-direction'' 
if there exists a path inside the set
$\PPP\cap\KKK(\XX,L)$ connecting those two faces
of $\SSS$ that are vertical to the $x$-axis.
Similarly for the other directions.
Thus $\Lambda_3=1$ indicates that the cell can be traversed along 
all 3 directions, while $\Lambda_c=1$ indicates that there exists
at least one direction along which the block is percolating.
$\Lambda_0=1$ indicates a blocking cell.

The local percolation probability in the ``$\alpha$''-direction
is now defined through
\begin{equation}
\lambda_\alpha(\phi,L) = \frac{
\sum_\XX \Lambda_\alpha(\XX,L)\delta_{\phi\phi(\XX,L)}}
{\sum_\XX\delta_{\phi\phi(\XX,L)}} .
\label{lpp2}
\end{equation}
The local percolation probability 
$\lambda_\alpha(\phi,L)$ gives the fraction of measurement
cells of sidelength $L$ with local porosity $\phi$ that are 
percolating in the ``$\alpha$''-direction.

\subsection{Total Fraction of Percolating Cells}
The total fraction of all cells percolating along the
``$\alpha$''-direction is given by integration 
over all local porosities as
\begin{equation}
p_\alpha(L)=\int_0^1\mu(\phi,L)\lambda_\alpha(\phi,L)\;d\phi
\label{pL1}
\end{equation}
This quantitiy provides an important characteristic for
network models.
For a network model it gives the fraction of network
elements (bond, sites etc.) which have to be permeable.

\section{Samples and Algorithms}
The data sets of three different sandstones are used in 
the analysis below.
Each data set consists of a threedimensional array of 0's 
and 1's indicating pore space $\PPP$ or matrix $\MMM$.
The array dimensions are $M_1,M_2$ and $M_3$.
The pore space $\PPP$ of the three samples are displayed
in Figures \ref{BereaA}, \ref{BrentB} and \ref{Sst20dC}.
Note that the representations are not to scale because
the resolution of each image is different.
The data were obtained by computerized microtomography
\cite{bakke}.

Table II gives a synopsis of the characteristics 
of the three samples which have been analysed.
\begin{center}
{\small TABLE II: Overview over properties of the data sets
for three reservoir sandstones.}\\[8pt]
\begin{tabular*}{\textwidth}{|l@{\extracolsep{\fill}}llllll|} \hline
Sample & Description & $a$ & $M_1\times M_2\times M_3$ & $L^*$ &
$\phb$ & $\kb$ \\\hline
A & Berea & $10\mu$m & $128\times 128\times 128$ & $260\mu$m &
$0.1775$ & $1100$mD \\
B & Brent & $2.7\mu$m & $180\times 217\times 217$ & $108\mu$m &
$0.1602$ & $470$mD \\
C & Sst20d & $30\mu$m & $73\times 128\times 128$ & $540\mu$m &
$0.2470$ & $20000$mD \\
\hline
\end{tabular*}
\end{center}
Here $a$ is the resolution, and $M_i$ are the dimensionless sidelengths
of the sample in units of $a$.
The bulk porosity was defined in eq. (\ref{bulk}) and the length
$L^*$ in eq. (\ref{lpd6}).
The permeability $\kb$, given in millidarcy, 
is the experimentally determined permeability 
of the sample from which the data sets were 
obtained.

The calculation of $\mu(\phi,L)$ is straightforward, and
proceeds exactly according to eq. (\ref{lpd2}).
Several possibilities exist for the choice of $\XX$
in $\KKK(\XX,L)$.
Originally \cite{hil91d} it was proposed to choose 
for $\XX$ a cubic lattice with lattice constant $L$ such that
$\bigcup_\XX\KKK(\XX,L)=\SSS$ and such that
the resulting set of $\KKK(\XX,L)$ are nonoverlapping
i.e. $\KKK(\XX,L)\cap\KKK(\XX^\prime,L)=\emptyset$
for $\XX\neq\XX^\prime$.
For the small data sets available this leads
to poor statistics with strong fluctuations in all results.
Therefore we use here a cubic lattice with smaller
lattice constants giving rise to overlapping cells.
The results below were obtained by using unit lattice
constant. In other words we used  for
$\XX$ all lattice sites except those whose
distance from the sample boundary is less than $L/2$.
It must be noted, however, that this method of positioning 
the cells gives progressively higher weight to the central
region of the sample.
This can, for large $L$, lead to small differences 
(roughly 0.005 in the present case)
between the bulk porosity $\phb$ as defined in eq. (\ref{bulk}) and
expected local porosity defined as $\int \phi\mu(\phi)d\phi$. 

The determination of $\Lambda_\alpha(\XX,L)$, i.e. of 
whether or not a cell is percolating
in a given direction, was carried out according to the 
well known Hoshen-Kopelman algorithm \cite{SA92}.
The choice of $\XX$ was the same as in the measurement
of $\mu$.
$\lambda_\alpha(\phi,L)$ was then calculated from eq. (\ref{lpp2})

\section{Results}
The first sample is Berea sandstone whose pore
space is displayed in Figure \ref{BereaA}.
In this case the resolution is $a=10\mu$m, and the
sidelengths of the sample are $M_1=M_2=M_3=128$.
Figure \ref{LPA} shows the 
local porosity distributions $\mu(\phi,L)$ 
with $L=40,80,120,300\mu$m exhibiting the
typical crossover between the limits
$L=0$ and $L=\infty$.
The curves are shown as dotted lines and marked by four 
different symbols corresponding to the four values of
$L$ as indicated in the legend.
Next, in the same Figure \ref{LPA}, the local percolation
probabilities $\lambda_3(\phi,L)$ 
are displayed for the same four values of $L$ that
were used for $\mu$.
The curves for $\lambda_3(\phi,L)$ are distinguished
from those for $\mu(\phi,L)$ by a solid line style.
The symbols used to indicate $L$ are the same in
both cases.
The ordinate for the $\lambda_3$-graphs is the right
axis, those for $\mu$-graphs is the left axis.
The local percolation probabilities $\lambda_3$ are
increasing from zero to one.
This expresses the fact that the full sample is
connected.
For $L=40\mu$m$=4a$ the sidelength of the measurement
cell corresponds to four voxels.
To have a conducting path in all three directions 
one needs at least 10 voxels of pore space.
This amounts to a porosity of roughly 0.16, and
hence all curves $\lambda_3(\phi,4a)$ must vanish
below $\phi\approx 0.16$.
Similarly to disconnect at least one of the three
directions one needs at least 16 voxels filling a
plane. 
Hence the curves $\lambda_3(\phi,4a)$ must equal
unity above $\phi\approx 1-0.25=0.75$.
This can be observed in Figure \ref{LPA} for Berea
and in Figures \ref{LPB} and \ref{LPC} for the
other samples.
For general $L$ and dimension $d$ the same consideration gives
that $\lambda_3$ vanishes below $(dL-d+1)/L^d$ and equals unity
above $1-1/L$.

It is instructive to compare $\mu$ with
$\lambda$ at a fixed $L$ by superposing them in the same plot. 
Such a plot is shown in Figure \ref{PDPPA} for $L=L^*$.
The characteristic length $L^*$ was defined in (\ref{lpd6}).
For sample A (Berea) its value is found
to be $L^*=260\mu$m.
To facilitate comparison the local porosity 
distribution $\mu(\phi,L^*)$ has been
rescaled such that its maximum equals unity.
All six local percolation functions $\lambda_\alpha$ are 
displayed in Figure \ref{PDPPA}.
The sample appears to be isotropic because
the three functions $\lambda_x,\lambda_y,\lambda_z$
all fall on top of each other.
The curves $\lambda_c$ and $\lambda_3$ are upper and
lower bounds for the region inside which the connectivity
increases from blocking to fully connected.
Note also that this band is shifted to the
left of the maximum of $\mu$ indicating that
Berea sandstone is well connected.
In fact the figure shows that an average cell (i.e.
a cell with local porosity around 0.18) is percolating
with probability larger than 0.75.
Only cells with local porosity much below average are 
blocking.

To investigate how heterogeneities in
the connectivity are reflected in the local percolation
probabilities, we have constructed an artificial modification
of the Berea sample.
To this end we have blocked roughly 1200 additional voxels
out of the total of $128^3\approx 2\cdot 10^6$ voxels
by blocking a plane of $1\times 100\times 100$
voxels inside the sample.
The orientation of the blocked plane was chosen
perpendicular to the $x$-direction, and the plane 
was centered in the $y$- and $z$-directions.
It was placed in the middel, i.e. into the 64th layer along
the $x$-direction.
This amounts to a small decrease of porosity by roughly
0.00058.
Note that the plane does not block the $x$-connectivity
completely, but leaves an open shell at the sample boundary.
The resulting modified porous microstructure
was visually indistinguishable from the unmodified
one from the perspective of Figure \ref{BereaA}.
Only when viewing the sample at right angles from 
the $y$- or $z$-direction it was possible to detect
a small modification.
In Figure \ref{PDPPABP} we display the same 
superposition of $\mu$ and $\lambda$ for the partially
modified sample that was shown in Figure \ref{PDPPA} for
the unmodified sample.
The local porosity distribution of the partially
$x$-blocked sample is almost identical to that
of the unmodified sample.
This remains true for all values of $L$,
and $L^*=260\mu$m is also unchanged.
The functions $\lambda_0,\lambda_c,\lambda_y$ and
$\lambda_z$ also remain almost unchanged.
$\lambda_x$ however differs from $\lambda_y$ and $\lambda_z$
as expected.
As a consequence also $\lambda_3$ falls significantly 
below the result of the unmodified sample.
The deviations of roughly 15\% give an order
of magnitude for the influence of connectivity
fluctuations on $\lambda_3$.
The difference between the results grows with 
increasing $L$.
This allows another important conclusion.
To characterize heterogeneities using local porosity
analysis it is necessary to measure both $\mu$ and $\lambda$
as functions of $L$ over a sufficiently wide range of $L$.
Choosing only one fixed $L$ may be misleading.

The dependence of the difference between the unmodified
Berea and the partially blocked sample on $L$ 
has been further quantified in Figure \ref{PLA}.
This figure shows the total fraction of percolating cells
determined according to eq. (\ref{pL1}).
The results for the original unmodified sample
are shown with solid lines, those for the
sample with a partially blocking plane
are shown as dotted lines.
This plot shows again that the unmodified sample 
is very isotropic because $p_x,p_y,p_z$ overlap.
In the partially blocked sample, however,
deviations start to appear around $L=150\mu$m
in $p_x(L)$ and $p_3(L)$ becoming more pronounced
at higher $L$.
Note that the modified sample shows a decrease
in $p_x(L)$ at $L\approx 300\mu$m while
all curves for the unmodified sample
are monotonously increasing.
Of course $p_x(L)$ must start to increase again 
at $L\gg 400\mu$m because the sample is still
connected on large scales.
Therefore one expects that nonmonotonous behaviour 
of $p(L)$ correlates with the length scale of 
heterogeneities in the connectivity.

The next sample is Brent sandstone, shown in Figure \ref{BrentB},
with a resolution of $a=2.7\mu$m, and sample dimensions
$M_1=180,M_2=217,M_3=217$.
Although the data set for this sample is the
largest one with respect to the number of
voxels, its absolute size is the smallest
of all samples.
As a consequence the statistics of this
sample is poor because it represents little more
than a few pores.
A larger sample seems necessary to obtain
a representative sampling of the pore space.

The local porosity distribution and local percolation probabilities
for this sample are shown in Figure \ref{LPB} using the same method
of plotting as in Figure \ref{LPA} (see above).
The superposition of $\mu(\phi,L^*)$ and $\lambda(\phi,L^*)$
is displayed in Figure \ref{PDPPB}.
Finally the total fraction of percolating cells is shown
in Figure \ref{PLB}.
This curve seems to indicate that, while the sample is isotropic
for small $L$, it shows increasing anisotropy at larger $L$.
This effect may, however, also be due to the poor statistics as
a result of the small absolute size of the system.

Sample C is a clean weakly consolidated sandstone of unkown origin
denoted below as {\em Sst20d} because its permeability is 20 Darcy.
The sample has
resolution $a=30\mu$m and sample dimensions
$M_1=73,M_2=128,M_3=128$, and its pore space is
displayed in Figure \ref{Sst20dC}.
The local porosity distribution for this sample is shown
in Figure \ref{LPC} together with local percolation probabilities.
Their superposition for $L=L^*$ is displayed in Figure \ref{PDPPC}.
Finally the total fraction of percolating cells is shown
in Figure \ref{PLC}.

Figures \ref{PDPPC} and \ref{PLC} indicate that
the sample is anisotropic because $\lambda_x$ is
significantly smaller than $\lambda_y$ and $\lambda_z$.
It has been checked that this is not a finite size
effect due to the sidelength in
the $x$-direction being shorter in this sample than,
say, in sample A.
This check was carried out by first dividing sample A 
in half along the $x$-direction, and then carrying out
the analysis on the remaining truncated sample.
This did not produce differences between 
$\lambda_x, \lambda_y$ and $\lambda_z$ for sample A.

Hence it must be concluded that sample C is 
anisotropic in its connectivity, being less
permeable in the $x$-direction than in the $y$-
and $z$-directions. 
We point out, however, that to the unaided eye
Figure \ref{Sst20dC} appears
visually isotropic.

\section{Comparison of Results and Discussion}
\begin{figure}[!ht] 
  \begin{center}\epsfig{figure=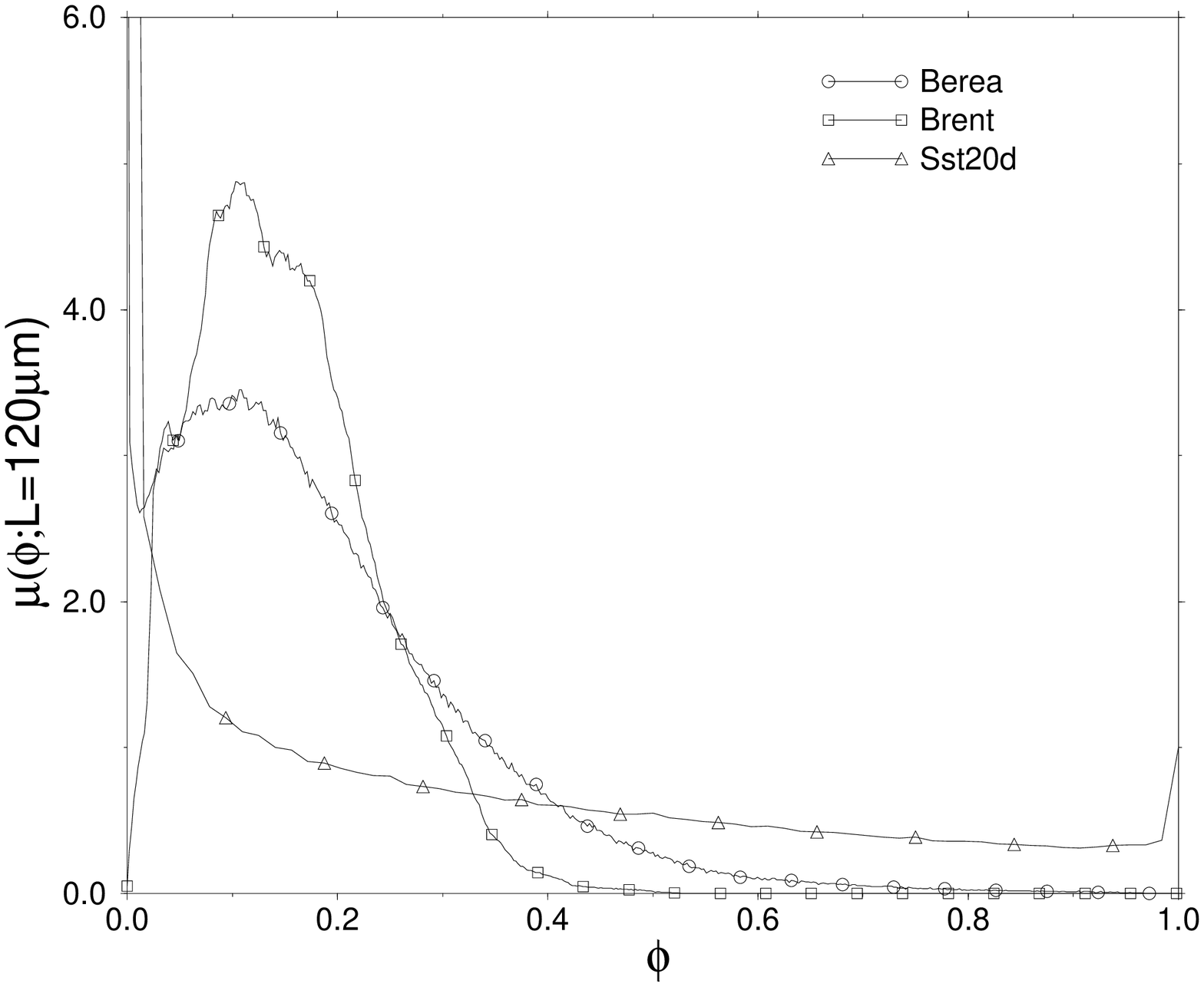,angle=-90,width=0.65\linewidth}\end{center}
  \caption{Local porosity distributions $\mu(\phi,L=120\mu$m$)$
    for all three samples}
  \label{LPD1}
\end{figure}

Having presented the results for the various samples
we now compare the samples against each
other.
Figure \ref{LPD1} shows the local porosity distribution
of all three samples at the same length $L=120\mu$m.
An important reason for the differences are different 
characteristic length scales for different samples.
Sample C has clearly the largest length scale because its
$\mu(\phi)$-curve is closest to the $L=0$ limit of eq. (\ref{lpd4}).
Next comes sample A, and sample B has the smallest
pores.
The difference in characteristic length scales may be 
eliminated by comparing the samples at some intrinsic length
scale such as the correlation length or the intrinsic
length $L^*$.

\begin{figure}[!ht] 
  \begin{center}\epsfig{figure=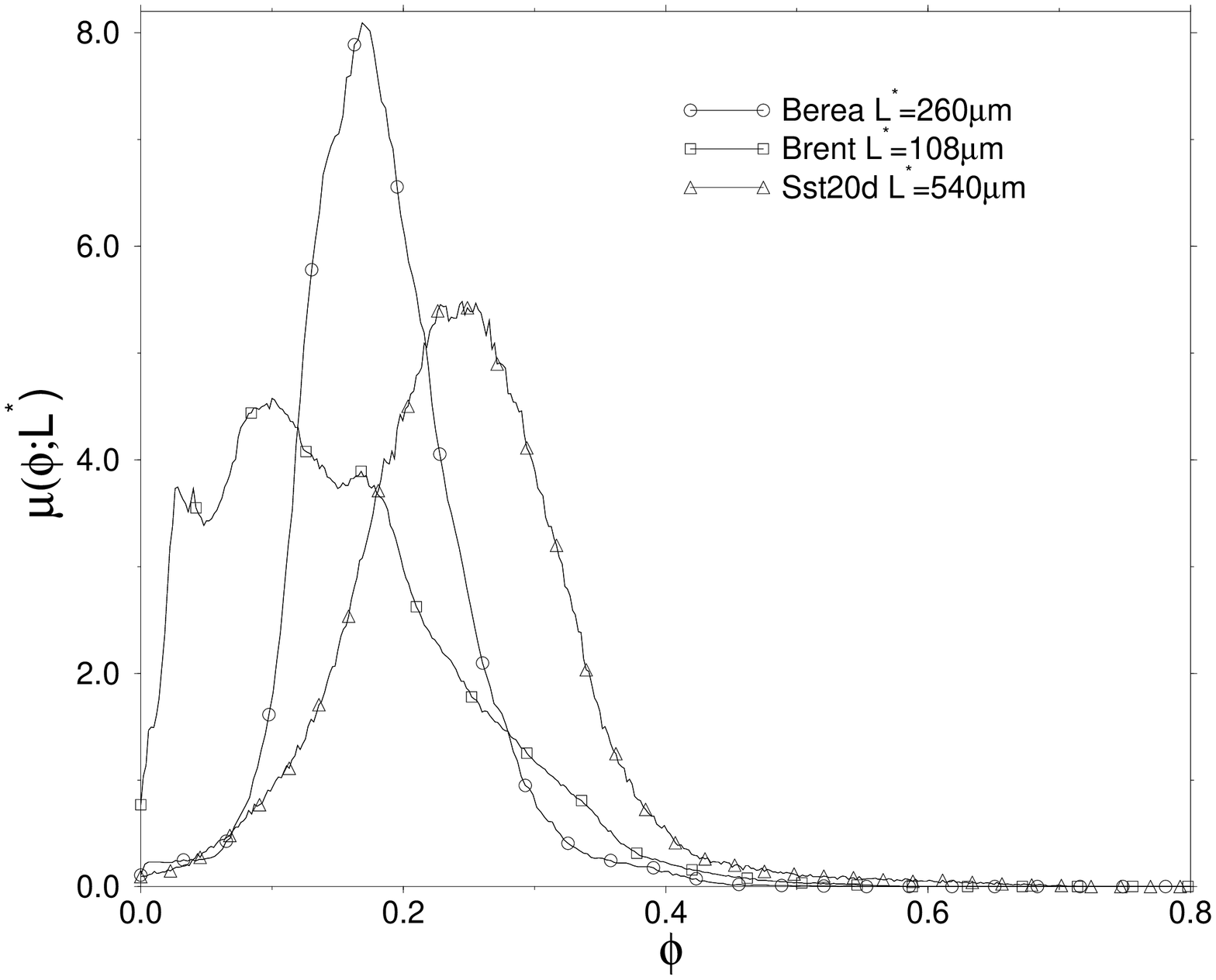,angle=-90,width=0.65\linewidth}\end{center}
  \caption{Local porosity distributions $\mu(\phi,L=L^*)$
    for all three samples}
  \label{LPD2}
\end{figure}

Figure \ref{LPD2} shows the comparison of
all three samples at the intrinsic length $L=L^*$.
Now the local porosity distributions resemble each 
other much more closely.
Nevertheless characteristic differences remain
not only in their peak position but also in their shape.
These may in part, but not entirely, be attributed to
the different porosities.
Sample A and sample B have nearly the same porosities,
but the shape of their $\mu(\phi,L^*)$
differs significantly.
The width of the curves indicates the strength of 
porosity fluctuations, and hence is a quantitative 
measure of heterogeneities in the porosity.
Using the width as a criterion we find that sample A 
is most homogeneous while sample B is most heterogeneous,
and sample C is intermediate.
This agrees with visual inspection of Figures
\ref{BereaA} through \ref{Sst20dC}, and
illustrates that $\mu(\phi)$ measures porosity
heterogeneities.
This fact was first demonstrated for twodimensional images
in \cite{hil92f} and suggests to use local porosity 
distributions and percolation probabilities as a quantitative 
method to compare microstructures between models and experiment.

In all three samples the behaviour of $\mu(\phi,L)$
as function of $L$ seems to approach the limits given in
eqs. (\ref{lpd4}) and (\ref{lpd5}).
This indicates that the samples approach macroscopic
homogeneity for $L\rightarrow\infty$ \cite{hil95d}. 
Of course much larger samples (particularly for sample B)
are needed to conclude this with certainty.

\begin{figure}[!ht] 
  \begin{center}\epsfig{figure=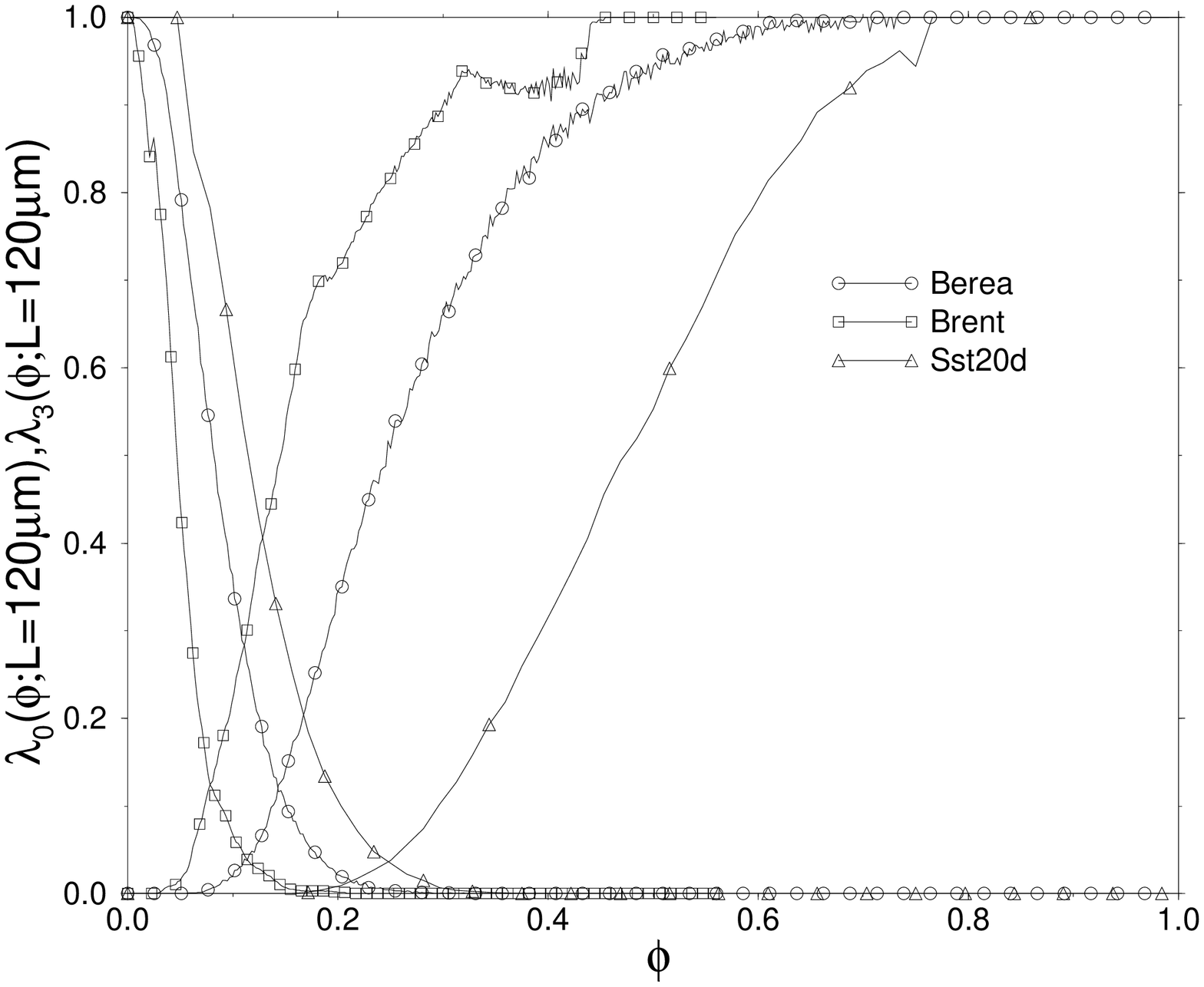,angle=-90,width=0.65\linewidth}\end{center}
  \caption{Local percolation probabilities $\lambda_3(\phi,L=120\mu$m$)$ and
    $\lambda_0(\phi,L=120\mu$m$)$ for all three samples}
  \label{LPP1}
\end{figure}

The behaviour of $\lambda(\phi,L)$ reflects the same
trend towards macroscopic homogeneity because these
functions approach a unit step at $\phb$
with increasing $L$.
The universality in the limit $L\rightarrow 0$
is reflected in the fact that the $\lambda$-curves
for small $L$ are very similar in Figures \ref{LPA},
\ref{LPB} and \ref{LPC}.
This was already discussed above for $L=4a$.
Of course, for fixed $L$, the range of $\phi$ over
which $\lambda$ changes coincides with the range of 
$\phi$ where $\mu(\phi,L)$ differs from zero.
Figure \ref{LPP1} compares the local percolation 
probabilities $\lambda_3$ and $\lambda_0$ for all 
samples at $L=120\mu$m.
Not surprisingly the three samples exhibit
very different behaviour analogous to the difference
in $\mu$ seen in Figure \ref{LPD1}.

\begin{figure}[!ht] 
  \begin{center}\epsfig{figure=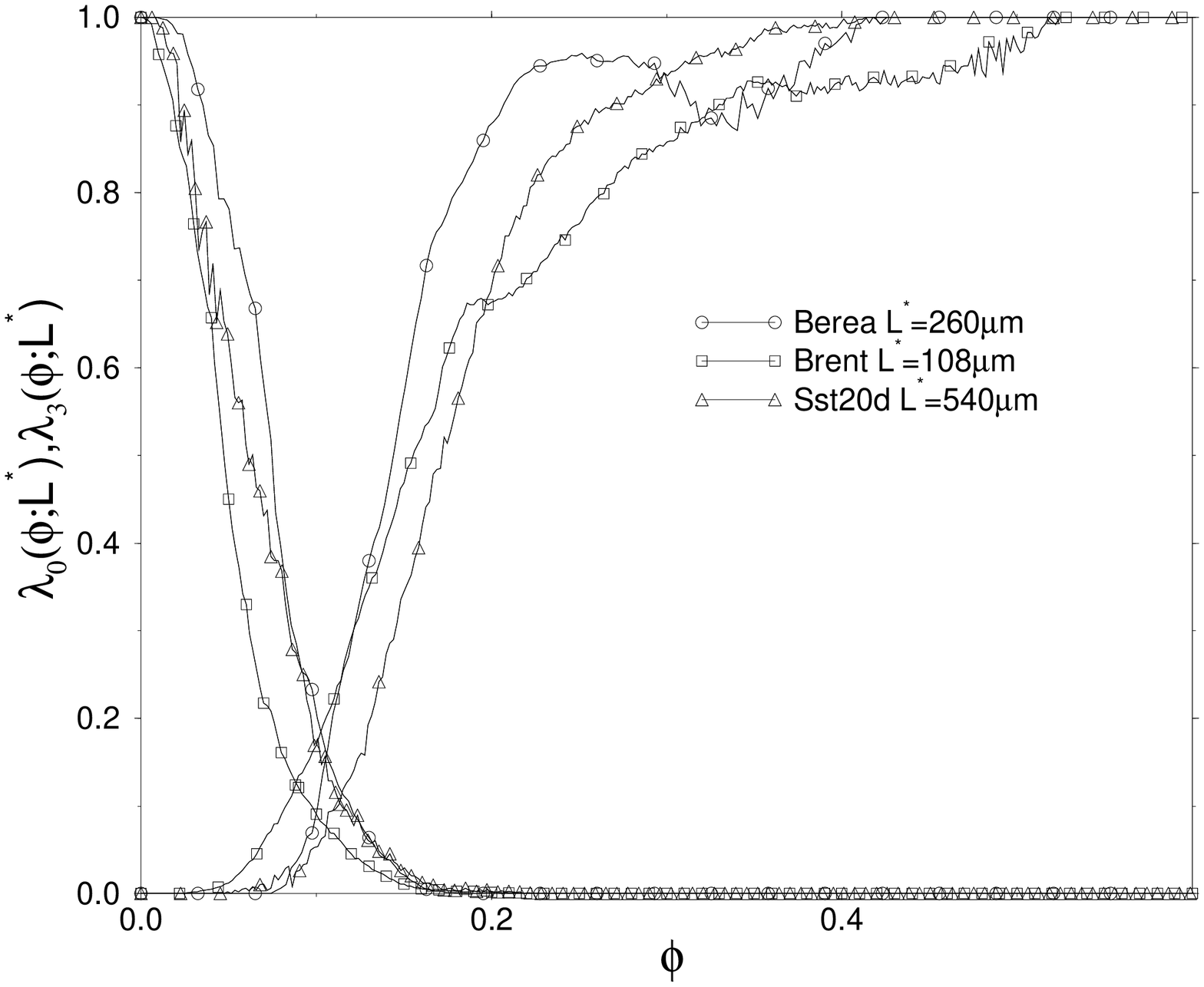,angle=-90,width=0.65\linewidth}\end{center}
  \caption{Local percolation probabilities $\lambda_3(\phi,L=L^*)$ and
    $\lambda_0(\phi,L=L^*)$
    for all three samples}
  \label{LPP2}
\end{figure}

Figure \ref{LPP2} shows $\lambda_3$ and $\lambda_0$
of all samples at the intrinsic length scale $L=L^*$.
Characteristic differences in shape appear which 
emphasize the different connectivity of the 
three samples.
These differences are not mere fluctuations because they are
of the same order of magnitude as the differences
introduced into $\lambda_3$ by the introduction
of a blocking plane into sample A that was discussed
above.
The curves are statistically most reliable
for $\lambda$ close to unity.
This can be seen from Figures \ref{PDPPA}, \ref{PDPPB}
and \ref{PDPPC}.
These plots show that the
maximum of $\mu$ occurs at $\phi$-values 
for which $\lambda$ is close to unity.
They also show that the five functions 
$\lambda_\alpha$ with $\alpha=c,x,y,z,3$ fall 
to the left of the maximum of $\mu$ for sample A and C,
while for sample B these functions change most rapidly
in the vicinity of the maximum of $\mu$.
Thus for sample A and C the change from blocking
to percolating occurs well below the average porosity
in the low porosity tail of $\mu$, indicating that
both samples have a very high degree of connectivity.
The modified sample A (with partially blocking plane)
shows a broader overlap between $\lambda_\alpha$ and
$\mu$ indicating lower connectivity (see Figure 
\ref{PDPPABP}), and sample B appears to have the 
lowest degree of connectivity.
In summary Figure \ref{LPP2} shows that although all the
samples are very well connected the fluctuations
in connectivity are different, and hence one
must also expect permeability fluctuations.

When comparing $\lambda_3$ for all samples it is seen
that this function reaches a plateau at large $\phi$.
For sample A there is a region around $\phi\approx 0.3$
where $\lambda_3$ decreases (see Fig \ref{PDPPA}).
This shows that local percolation probabilities 
are not always strictly monotonous (as might have
been expected) but may exhibit minima and maxima
indicating a variable fraction of blocking cells
and hence connectivity heterogeneities at 
intermediate scales.

\begin{figure}[!ht] 
  \begin{center}\epsfig{figure=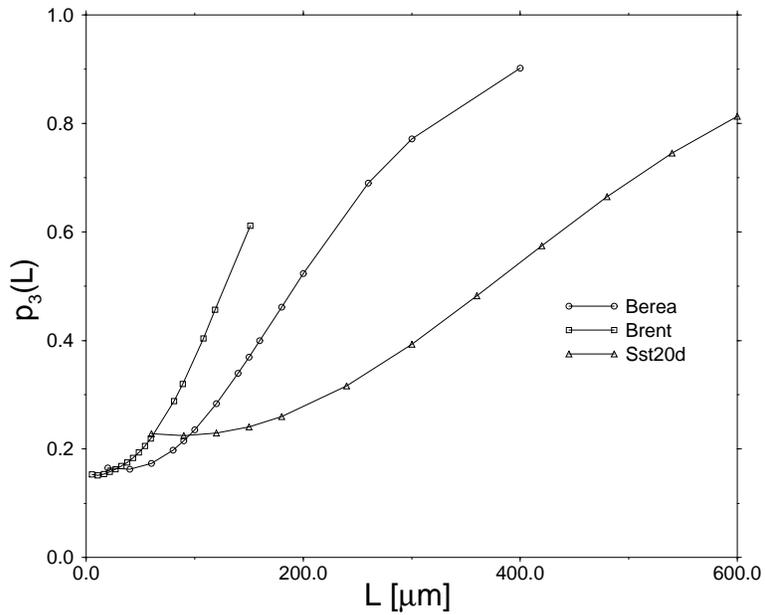,angle=-90,width=0.65\linewidth}\end{center}
  \caption{Total fraction of percolating cells $p_3(L)$ for all three samples.}
  \label{PL1}
\end{figure}

Finally it is instructive to compare the total fraction
of percolating cells $p(L)$ for all samples.
Figures \ref{PLA}, \ref{PLB} and \ref{PLC}
show that samples A and possibly B are
essentially isotropic while sample C has clearly
anisotropic connectivity.
Figure \ref{PL1} shows $p_3(L)$ as calculated
from eq. (\ref{pL1}) for all samples.
This plot can be used for constructing network models
in two ways.
Firstly, if each $\KKK(\XX,L)$ is used to represent
a site in the network model, $p_3(L)$  provides an
estimate for the fraction of percolating network units 
as a function of the networks lattice constant.
Secondly, $p_3(L)$ exhibits intrinsic length scales.
Extrapolating a tangent at the inflection point 
of each curve to $p=1$ gives
a length scale which could be interpreted as
the minimum length scale for a representative
elementary volume (REV) needed in homogenization
and other averaging procedures \cite{BB90,hil95d}.
In this way one finds approximately
350$\mu$m for sample A, 230$\mu$m for sample B
and 700$\mu$m for sample C as the smallest
sidelength for an REV.

\begin{figure}[!ht]
  \begin{center}\epsfig{figure=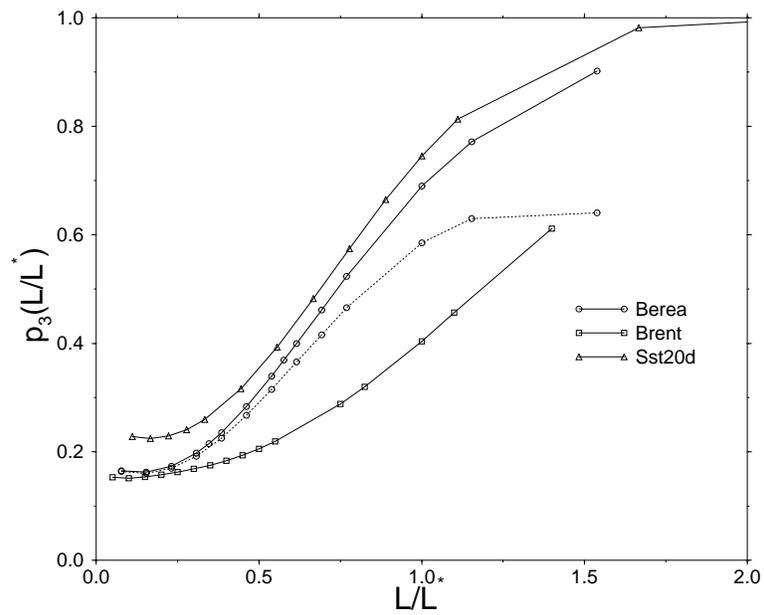,angle=-90,width=0.65\linewidth}\end{center}
  \caption{Total fraction of percolating cells $p_3(L)$ for all three samples
    rescaled with $L^*$.
    The dotted line with circles is the result for the Berea sandstone
    (sample A) with a partially blocking $yz$-plane inside it.}
  \label{PL2}
\end{figure}

The results presented here suggest
that none of the three samples can be adequately modelled by
a critical site percolation network.
The reason is that in such a model the fraction of
percolating sites should equal $p_3$ and hence would
be strongly $L$-dependent, except for very small or
very large $L$.
Hence, plots such as Figures \ref{PL1} or \ref{PL2} provide
information on how to choose the network elements 
(site, bonds) of a network model and how to relate
the length scale of the real rock to the networks
lattice constant.

Figure \ref{PL2} shows $p_3(L/L^*)$ for all samples
as solid lines. 
The dotted line corresponds to the modified sample A with
a partially blocking plane.
The samples are again significantly different.
The vertical shift is in part due to differences in
porosity because generally $p_3(L=0)=\phb$.
The sample with the blocking plane indicates that
it is possible to have intermediate plateaus
and nonmonotonicity, at least if large scale
heterogeneities are present.
The unmodified homogeneous samples exhibit different widths over 
which the curves increase from $\phb$ to unity.
These widths may be used as a quantitative measure of
fluctuations in connectivity, both in theoretical
models and in experiment.

ACKNOWLEDGEMENT:
The authors are grateful to Dr.P.E. {\O}ren (Statoil), Dr.S. Bakke
(Statoil), Dr.R. Hazlett (Mobil Exploration and Production) and
Brookhaven National Laboratories for providing the sample
data sets, and to the
Deutsche Forschungsgemeinschaft for financial support.

\bibliographystyle{ieeetr}
\textheight24.5cm

\newpage
\section*{Figures}

\begin{figure}[!h] 
  \caption{Threedimensional pore space of
    Berea sandstone (sample A).
    The resolution is $a=10\mu$m, the sample dimensions are
    $M_1 = 128$,
    $M_2 = 128$,
    $M_3 = 128$.
    The bulk porosity is $\phb=0.1775$.
    The pore space is indicated in blue, the matrix space is
    transparent.}
  \label{BereaA}
\end{figure}

\begin{figure}[!h]
  \caption{Threedimensional pore space reconstruction of
    Brent sandstone (sample B).
    The resolution is $a=2.7\mu$m, the sample dimensions are
    $M_1 = 180$,
    $M_2 = 217$,
    $M_3 = 217$.
    The bulk porosity is $\phb=0.1602$.
    The pore space is indicated in blue, the matrix space is
    transparent.}
  \label{BrentB}
\end{figure}

\begin{figure}[!h]
  \caption{Threedimensional pore space reconstruction of a weakly
    consolidated sandstone (sample C).
    The resolution is $a=30\mu$m, the sample dimensions are
    $M_1 = 73$,
    $M_2 = 128$,
    $M_3 = 128$.
    The bulk porosity is $\phb=0.2450$.
    The pore space is indicated in blue, the matrix space is
    transparent.}
  \label{Sst20dC}
\end{figure}

\begin{figure}[!h]
  \begin{center}
    \epsfig{figure=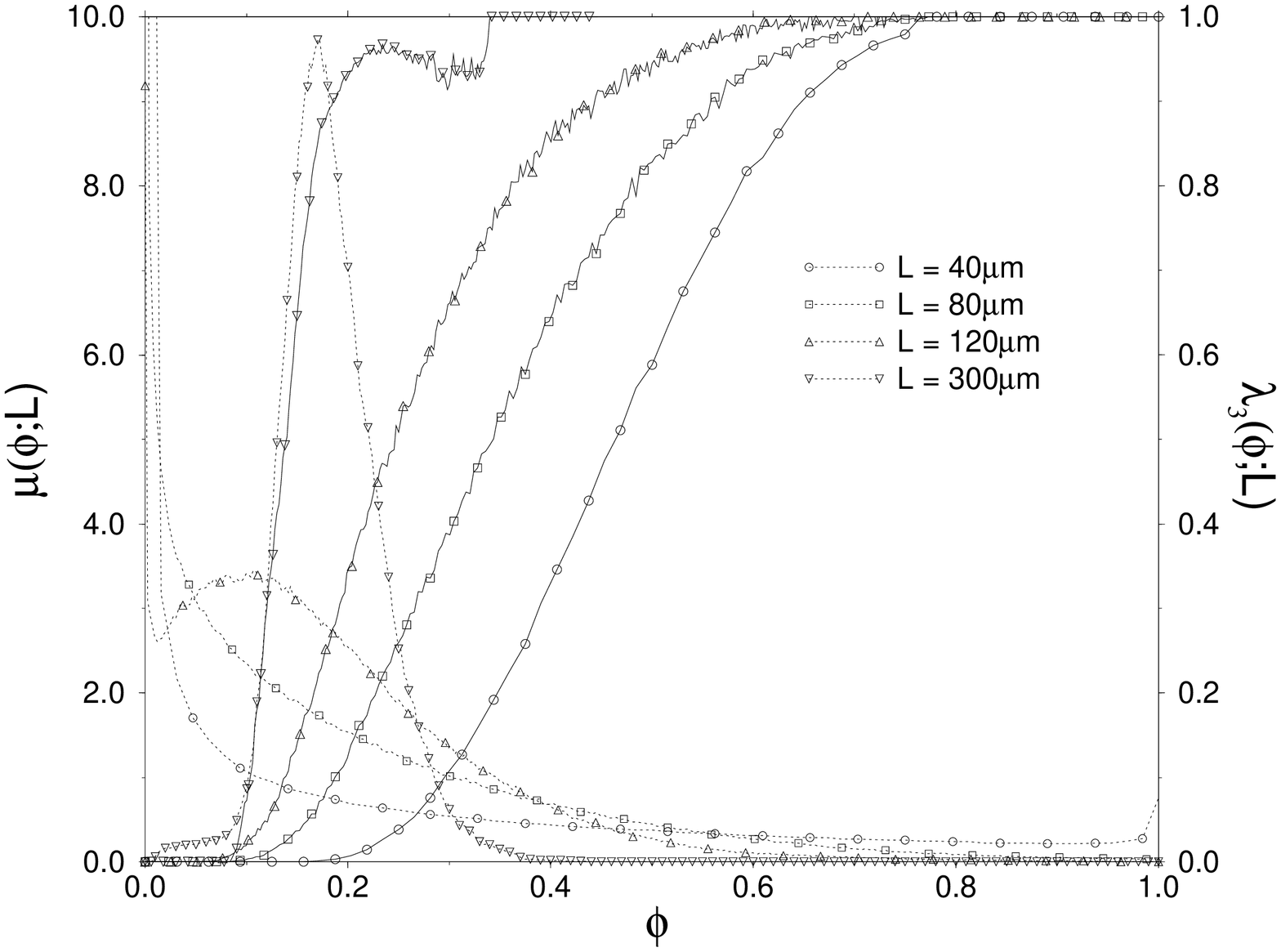,angle=-90,width=0.65\linewidth}
  \end{center}
  \caption{Local porosity distributions $\mu(\phi,L)$ (dotted lines) and
    local percolation probabilities $\lambda_3(\phi,L)$ (solid lines)
    for Berea sandstone (sample A, cf. Figure \ref{BereaA}).
    Four different values for $L$ are indicated by different
    symbols defined in the legend.
    The ordinate for the graphs of $\mu(\phi,L)$ is on the left, 
    the ordinate for $\lambda_3(\phi,L)$ is on the right as 
    indicated by the axis labels.}
  \label{LPA}
\end{figure}

\clearpage

\begin{figure}[!ht]
  \begin{center}\epsfig{figure=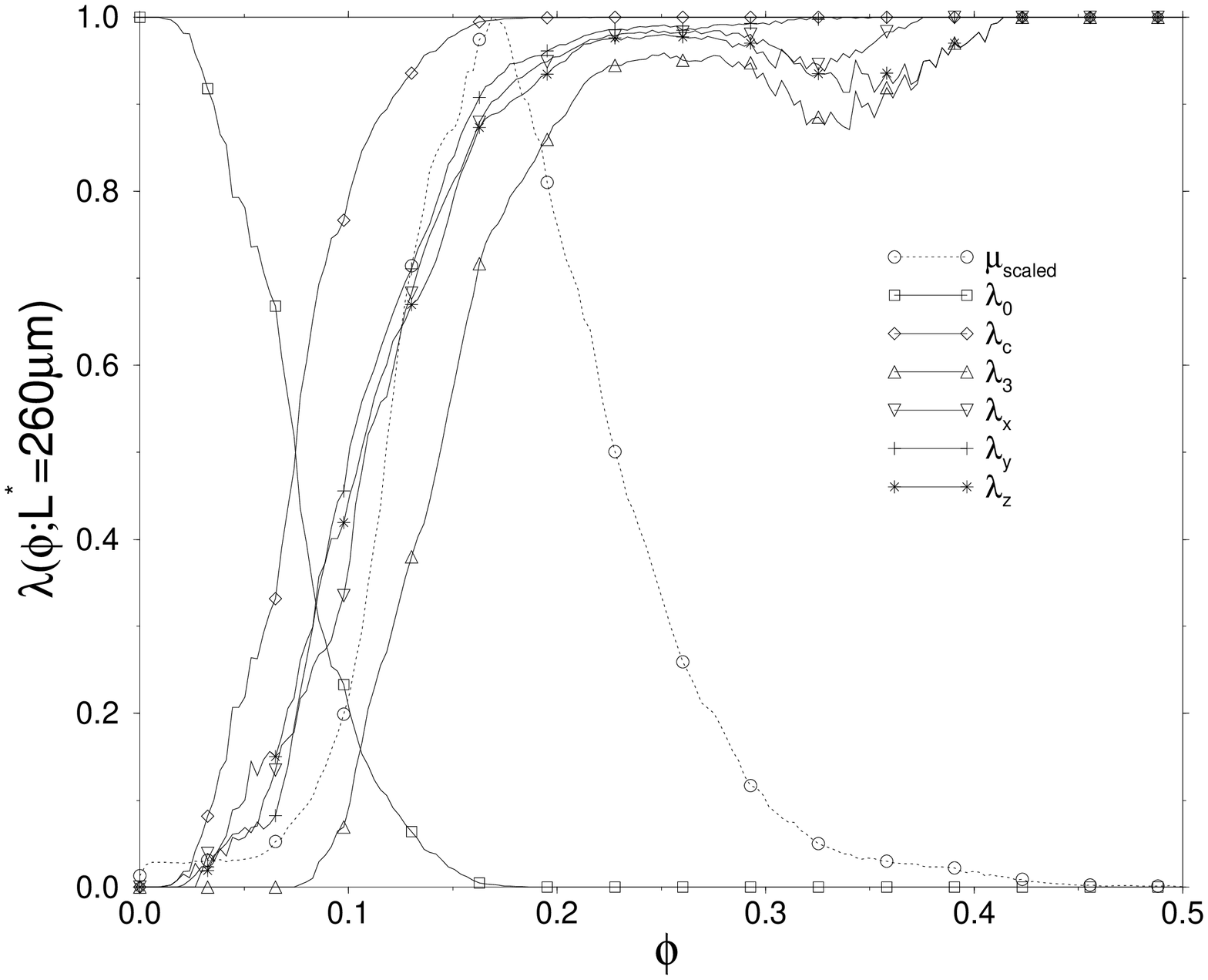,angle=-90,width=0.65\linewidth}\end{center}
  \caption{Local percolation probabilities $\lambda_\alpha(\phi,L^*)$ 
    with $\alpha=0,3,c,x,y,z$ and $L^*=260\mu$m
    for Berea sandstone (sample A) shown in Figure \ref{BereaA}.
    The dotted curve with circular symbols is the 
    local porosity distribution at $L=L^*=260\mu$m
    rescaled to have maximum 1.}
  \label{PDPPA}
\end{figure}

\begin{figure}[!ht]
  \begin{center}\epsfig{figure=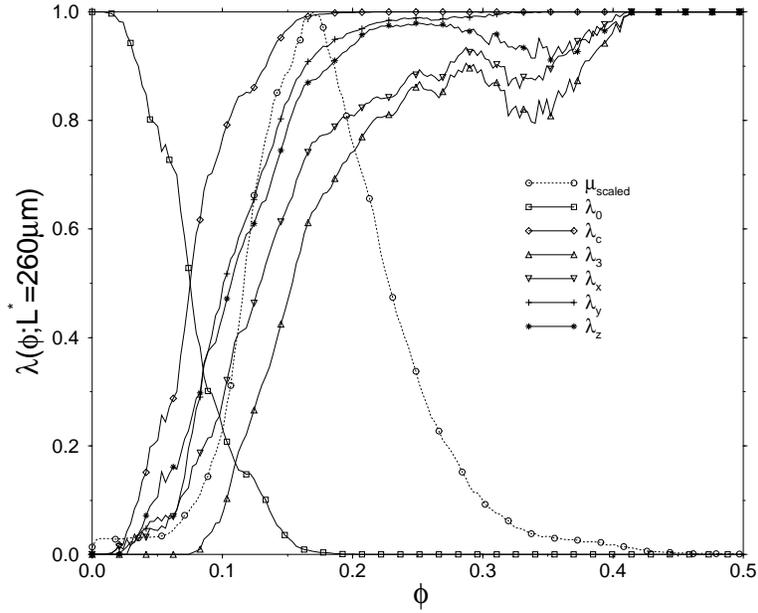,angle=-90,width=0.65\linewidth}\end{center}
  \caption{Same as Figure \ref{PDPPA} but for sample A modified
    with partially blocking plane in $x$-direction. See text for
    details.}
  \label{PDPPABP}
\end{figure}

\clearpage

\begin{figure}[!ht]
  \begin{center}\epsfig{figure=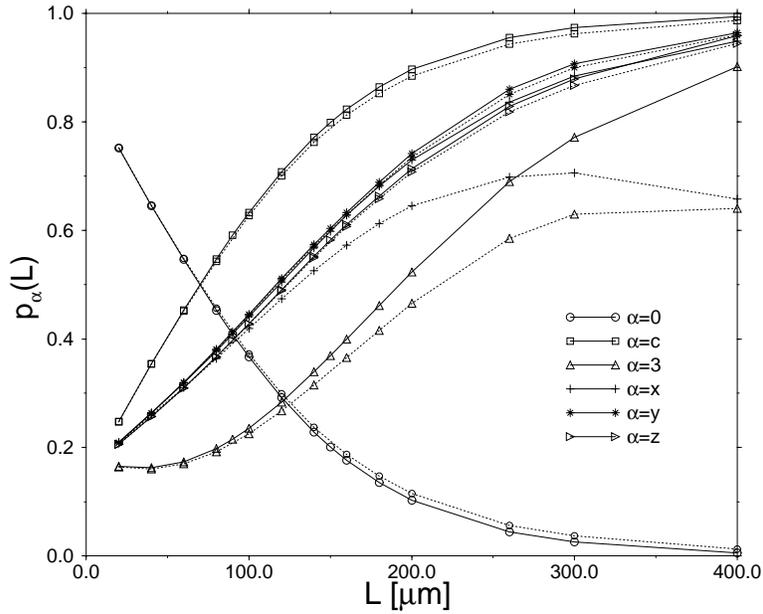,angle=-90,width=0.65\linewidth}\end{center}
  \caption{Total fraction of percolating cells $p_\alpha(L)$
    with $\alpha=0,3,c,x,y,z$
    for Berea sandstone (sample A, cf. Figure \ref{BereaA})
    shown as solid lines, and for the modified sample A
    (with partially blocking $yz$-plane) shown as dotted lines.}
  \label{PLA}
\end{figure}

\begin{figure}[!ht]
  \begin{center}\epsfig{figure=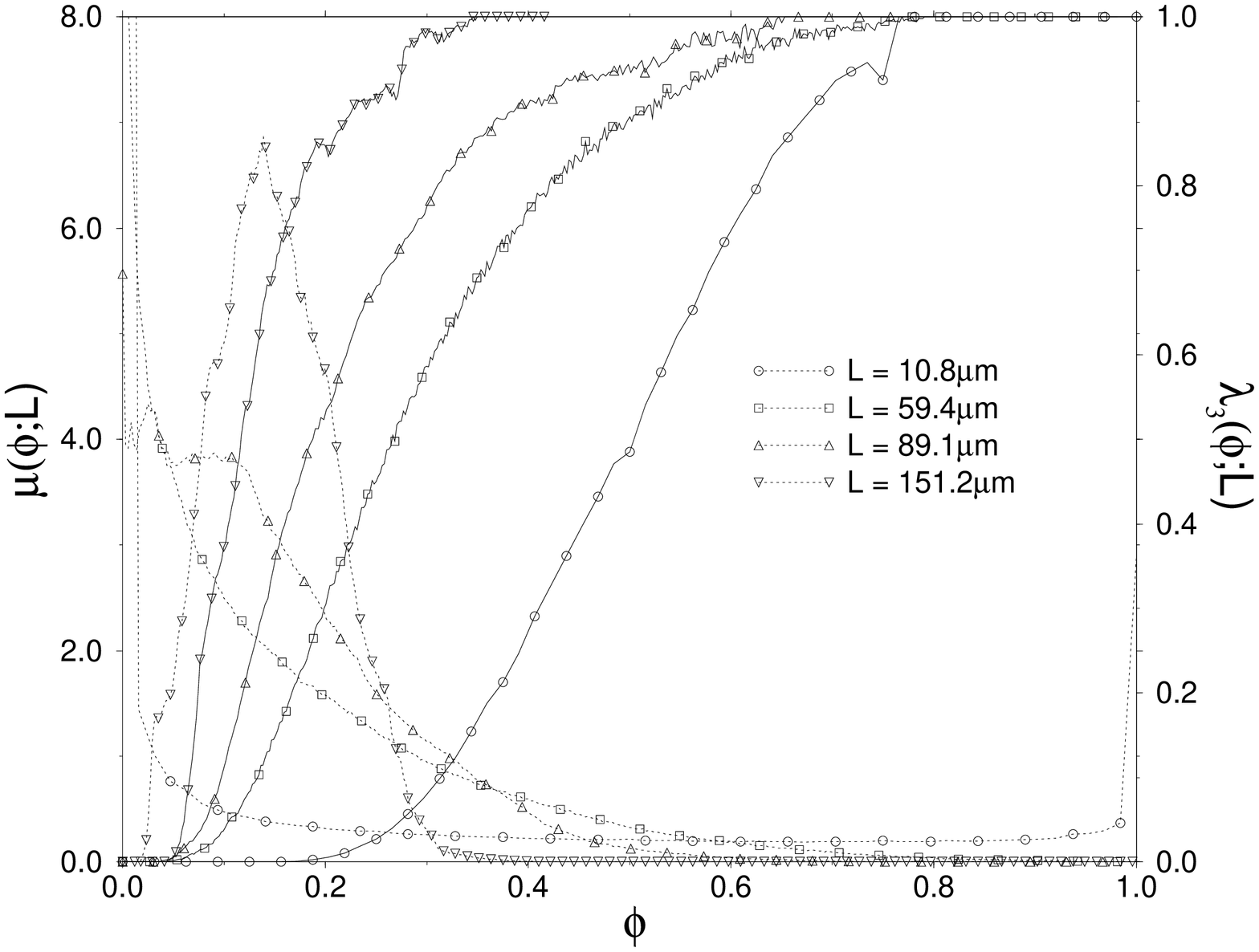,angle=-90,width=0.65\linewidth}\end{center}
  \caption{Local porosity distributions $\mu(\phi,L)$ (dotted lines) and
    local percolation probabilities $\lambda_3(\phi,L)$ (solid lines)
    for Brent sandstone (sample B) shown in Figure \ref{BrentB}.
    Four different values for $L$ are indicated by different
    symbols defined in the legend.
    The ordinate for the graphs of $\mu(\phi,L)$ is on the left, 
    the ordinate for $\lambda_3(\phi,L)$ is on the right as 
    indicated by the axis labels.}
  \label{LPB}
\end{figure}

\clearpage

\begin{figure}[!ht]
  \begin{center}\epsfig{figure=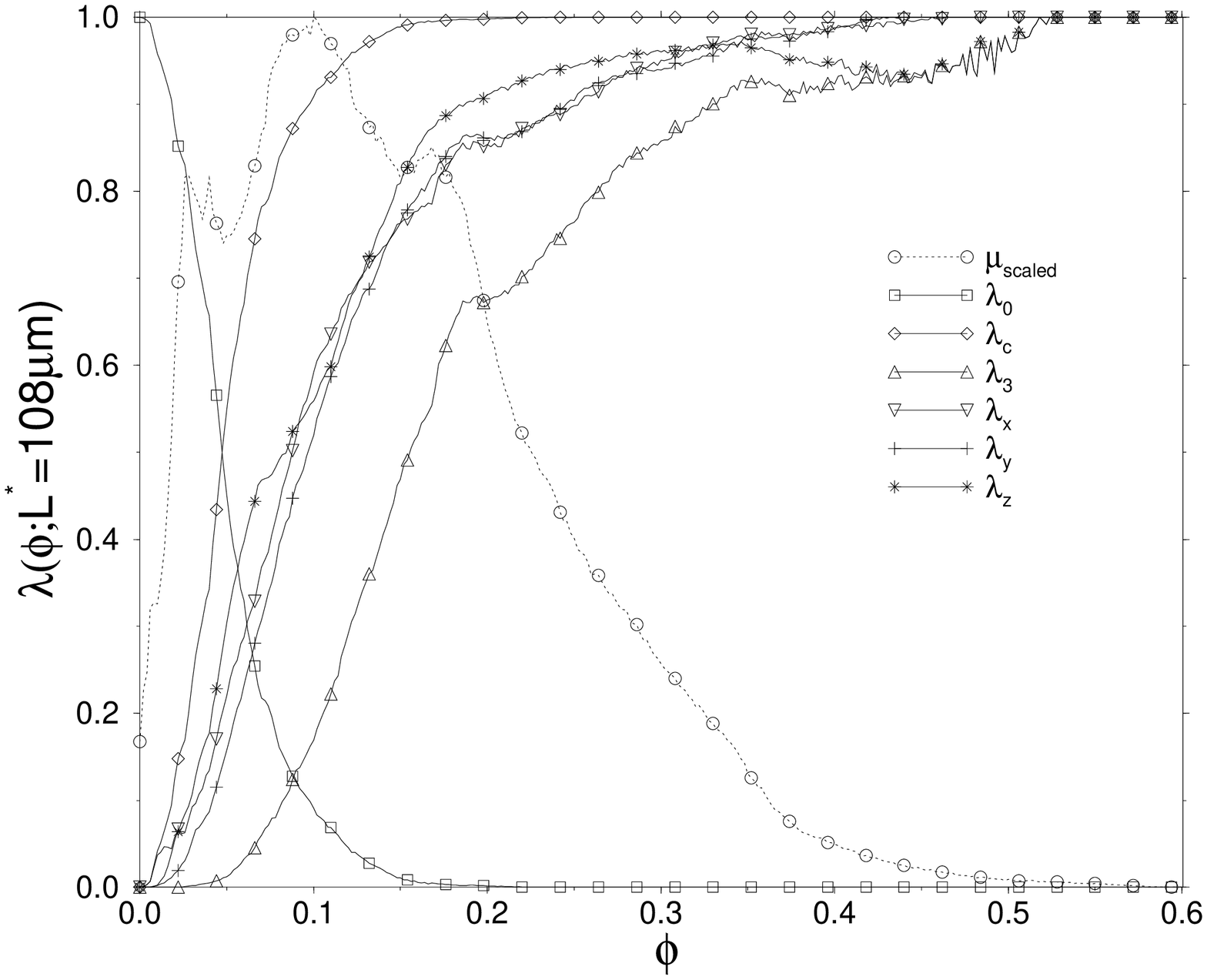,angle=-90,width=0.65\linewidth}\end{center}
  \caption{Local percolation probabilities $\lambda_\alpha(\phi,L^*)$ 
    with $\alpha=0,3,c,x,y,z$ and $L^*=108\mu$m
    for Brent sandstone (sample B) shown in Figure \ref{BrentB}.
    The dotted curve with circular symbols is the 
    local porosity distribution at $L=L^*=108\mu$m
    rescaled to have maximum 1.}
  \label{PDPPB}
\end{figure}

\begin{figure}[!ht]
  \begin{center}\epsfig{figure=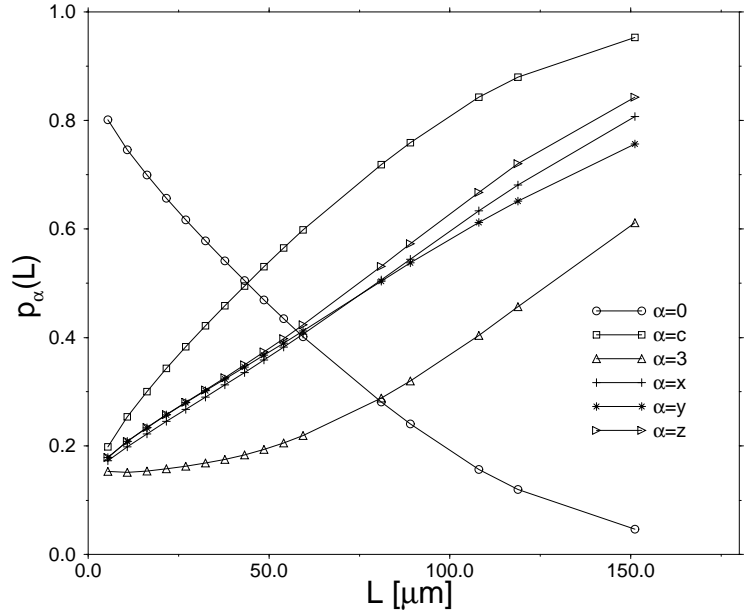,angle=-90,width=0.65\linewidth}\end{center}
  \caption{Total fraction of percolating cells $p_\alpha(L)$
    with $\alpha=0,3,c,x,y,z$
    for Brent sandstone (sample B) shown in Figure \ref{BrentB}.}
  \label{PLB}
\end{figure}

\begin{figure}[!ht]
  \begin{center}\epsfig{figure=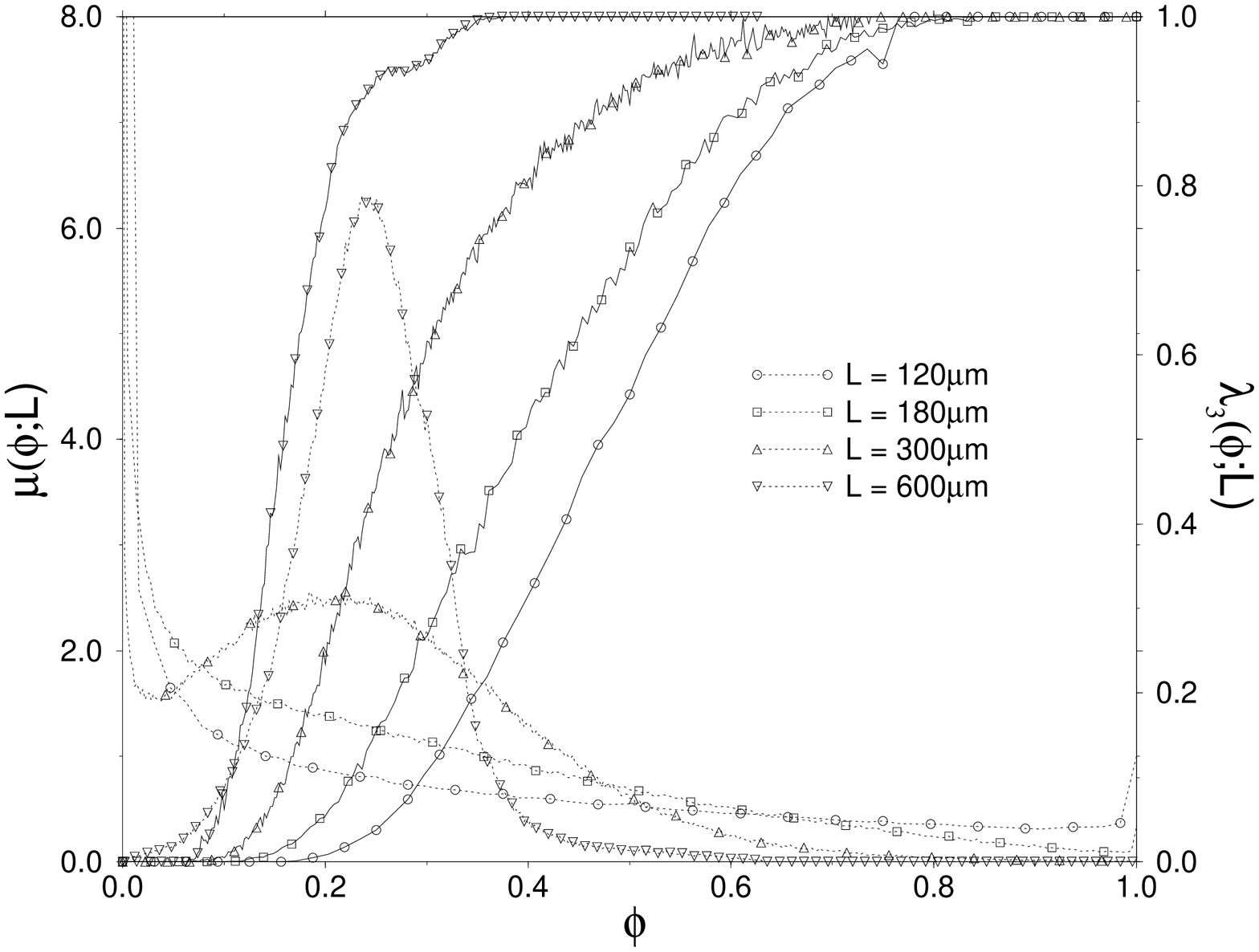,angle=-90,width=0.65\linewidth}\end{center}
  \caption{Local porosity distributions $\mu(\phi,L)$ (dotted lines) and
    local percolation probabilities $\lambda_3(\phi,L)$ (solid lines)
    for the sandstone (sample C) shown in Figure \ref{Sst20dC}.
    Four different values for $L$ are indicated by different
    symbols defined in the legend.
    The ordinate for the graphs of $\mu(\phi,L)$ is on the left, 
    the ordinate for $\lambda_3(\phi,L)$ is on the right as 
    indicated by the axis labels.}
  \label{LPC}
\end{figure}

\begin{figure}[!ht]
  \begin{center}\epsfig{figure=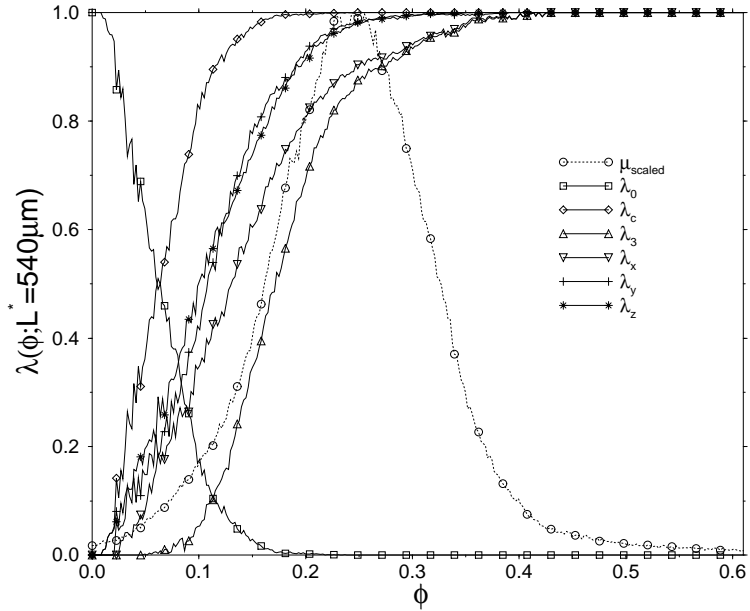,angle=-90,width=0.65\linewidth}\end{center}
  \caption{Local percolation probabilities $\lambda_\alpha(\phi,L)$ 
    superposed on local porosity distribution at $L=L^*=504\mu$m
    for the sandstone shown in Figure \ref{Sst20dC}.}
  \label{PDPPC}
\end{figure}

\begin{figure}[!ht] 
  \begin{center}\epsfig{figure=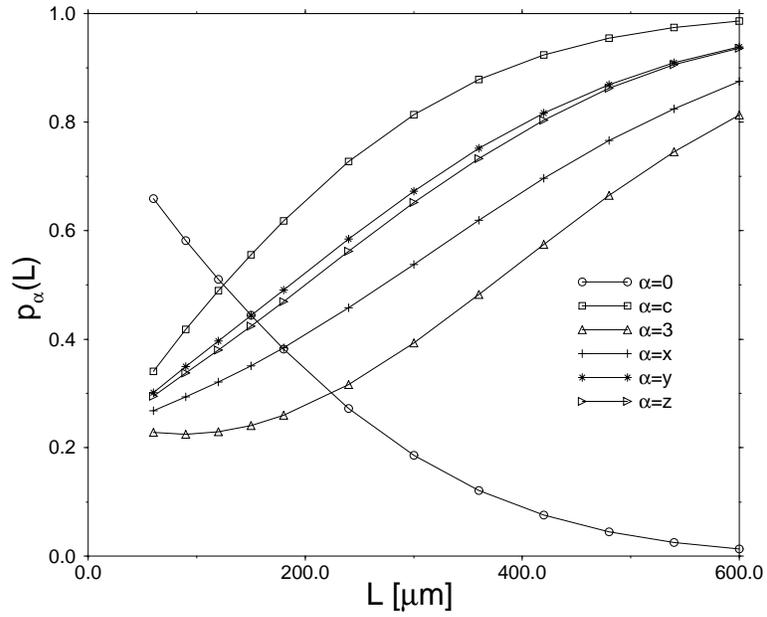,angle=-90,width=0.65\linewidth}\end{center}
  \caption{Total fraction of percolating cells $p_\alpha(L)$
    for the sandstone shown in Figure \ref{Sst20dC}.}
  \label{PLC}
\end{figure}

\end{document}